# Analytical Study of Surface Plasmon-Phonon Polaritons in Nonlinear-Graphene-LiF Heterostructures in the far-infrared region

**Mohammad Bagher Heydari** [1,*], **Ali Abdollahi** [1], **Sina Asgari** [2]

[1,*] Faculty of Technical and Engineering, Imam Khomeini International University, Qazvin, Iran
[2] Department of Electrical Engineering, University of Zanjan, Zanjan, Iran

[*] Corresponding author: mb.heydari@eng.ikiu.ac.ir

**Abstract:** In this paper, a new heterostructure based on the hybridization of graphene-LiF layers with a nonlinear material is introduced and studied. An analytical model is obtained by starting on Maxwell's equations and applying boundary conditions. The numerical results are depicted and discussed in detail. A high value of FOM (FOM=24.5) at the frequency of 9.22 THz is reported for the chemical potential of $\mu_c = 0.2\ ev$. Our results show that the propagation features of the proposed structure can be varied by the graphene parameters and the nonlinearity inside and outside the phononic band. The Hybridization of graphene with a nonlinear medium and a polar dielectric like LiF can support high levels of confinement with low optical loss, which makes this platform a unique candidate for THz applications.

## 1. Introduction

Graphene, a two-dimensional structure consisting of carbon atoms in a honeycomb lattice, has been attracted immense interest among scientists due to its unique properties such as tunable conductivity [1] and large optical nonlinearity [2]. Between these extraordinary features, its ability to propagate tunable Surface Plasmon Polaritons (SPP) in the THz frequencies makes it a suitable platform for "Plasmonics" science with various devices such as couplers [3, 4], filters [5-7], waveguides [8-27], sensors [28, 29], and resonators [30, 31]. The tunability of SPPs on a graphene sheet can be obtained by changing its chemical potential via electrostatic bias or chemical doping. Moreover, graphene can also support anisotropic SPPs by applying a perpendicular magnetic bias to its surface, which can be utilized to design and fabricate non-reciprocal devices such as circulators [32, 33] and isolators [34, 35].

Polar dielectrics can support high levels of confinement with low optical loss, originating from their optical phonon interactions with the material. There are several polar dielectrics in the literature with exotic features such as Hexagonal-Boron-Nitride (hBN) [36, 37], SiC [38, 39], $CaF_2$ [40], and GaAs [41, 42]. In the far-infrared region, LiF is a good candidate for the excitation of Surface Phonon Polaritons (SPhP) [43, 44].

In the far-infrared frequencies, graphene has shown some exotic properties [45, 46]. Hybridization of graphene and a polar dielectric can support Hybrid Surface Plasmon Phonon Polaritons (HSP[3]) [47, 48] which has potential applications such as plasmon-induced transparency [49], and planar focusing [50]. Moreover, the combination of graphene with a nonlinear medium enhances the propagating features of SPPs such as Figure of Merit (FOM) [51-56]. For instance, a high value of FOM is reported in [51], where a Kerr-type material is used in a graphene-based device. Therefore, hybridization of graphene-LiF heterostructure with a nonlinear material can improve the FOM of the structure, improves the localization of HSP[3], and also can give the designer high levels of freedom to adjust and change the propagating characteristics of such tunable devices.

To the best of our knowledge, no published work has reported a hybrid graphene-LiF-Nonlinear heterostructure in the far infrared frequencies. In this paper, after introducing this new heterostructure, an analytical model will be proposed by starting Maxwell's equations and solving the wave equation inside each layer. Then, by applying the boundary conditions at the borders, a mathematical dispersion relation will be derived for our nano-structure. In section 3, analytical results will be depicted and investigated. In this section, it has been shown that our proposed device is a tunable structure, in which its propagating properties can adjust via the chemical potential and the nonlinear coefficient. Finally, section 4 concludes the article.

## 2. The proposed structure and its analytical model

Fig. 1 represents the schematic of the proposed structure, where a graphene sheet is located on a LiF dielectric and the whole structure is placed on the SiO$_2$-Si layers. The cladding is assumed to be a nonlinear material with the following permittivity [57]:

$$\varepsilon_{NL} = \varepsilon_L + \alpha |\boldsymbol{E}|^2 \tag{1}$$

In (1), $\varepsilon_L$ is the linear part of permittivity and $\alpha$ is the nonlinear coefficient. One can write relation (1) as follows:

$$\varepsilon_{NL} = \varepsilon_L + \alpha \eta^2 |\boldsymbol{H}|^2 = \varepsilon_L + \alpha \frac{\mu_0}{\varepsilon_L} |\boldsymbol{H}|^2 = \varepsilon_L + \alpha \frac{1}{\varepsilon_0 \varepsilon_L c^2} |\boldsymbol{H}|^2 \tag{2}$$

Where $c, \varepsilon_0, \mu_0, \eta$ are the velocity of light, the permittivity, the permeability, and the impedance of free space, respectively. In the literature, Graphene conductivity is given by familiar relation, called Kubo's formula [1]:

$$\sigma(\omega, \mu_g, \Gamma, T) = \frac{-je^2}{4\pi\hbar} \ln\left[\frac{2|\mu_g| - (\omega - j2\Gamma)\hbar}{2|\mu_g| + (\omega - j2\Gamma)\hbar}\right] + \frac{-je^2 K_B T}{\pi\hbar^2(\omega - j2\Gamma)} \left[\frac{\mu_g}{K_B T} + 2\ln\left(1 + e^{-\mu_g/K_B T}\right)\right] \tag{3}$$

In (3), $\Gamma$ is the scattering rate, $T$ is the temperature, and $\mu_g$ is the chemical potential of graphene. Furthermore, $\hbar$ is the reduced Planck's constant, $K_B$ is Boltzmann's constant, ω is radian frequency, and $e$ is the electron charge.

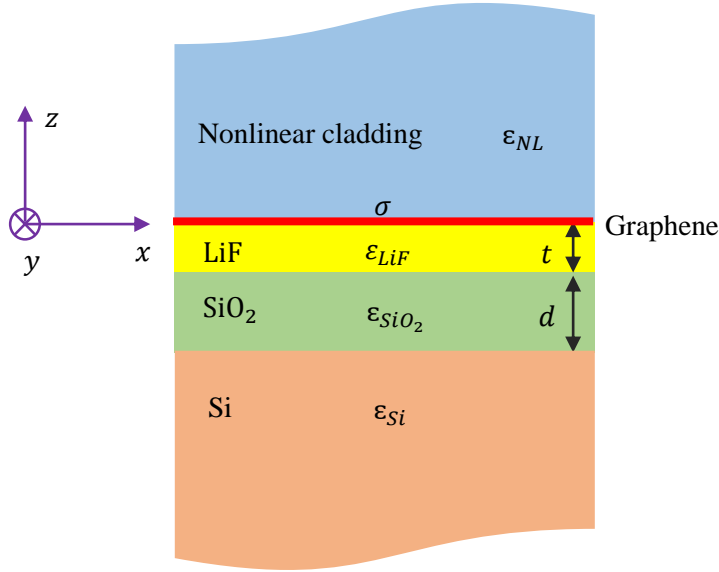

**Fig. 1.** The schematic of the proposed structure.

Here, we study TM mode with the following components: $E_x, E_z, H_y$. The propagation of HSP$^3$ is in the x-direction ($e^{j\beta x}$). So, the relation (1) can be written as:

$$\varepsilon_{NL} = \varepsilon_L + \alpha' |H_y|^2 \tag{4}$$

Where

$$\alpha' = \frac{\alpha}{\varepsilon_0 \varepsilon_L c^2} \tag{5}$$

is a nonlinear coefficient defined in (4). Now, consider the Maxwell's equations inside the nonlinear material (suppose $e^{-j\omega t}$):

$$\nabla \times \boldsymbol{E} = j\omega\mu_0 \boldsymbol{H} \tag{6}$$

$$\nabla \times \boldsymbol{H} = -j\omega\varepsilon_0 \varepsilon_{NL} \boldsymbol{E} \tag{7}$$

As mentioned before, the TM propagating mode has the components of $E_x, E_z, H_y$. By substituting these in (6)-(7), we achieve to:

$$\frac{d^2 H_y}{dz^2} - \gamma_{NL}^2 H_y + k_0^2 \alpha' H_y^3 = 0 \tag{8}$$

In (8), $k_0$ is the free-space wavenumber and $\gamma_{NL}$ is the wave constant:

$$\gamma_{NL}^2 = \beta^2 - k_0^2 \varepsilon_L \tag{9}$$

The transverse component of electric fields in the nonlinear media are derived by the Maxwell's equations of (6)-(7):

$$E_x = \frac{1}{j\omega\varepsilon_0 \varepsilon_{NL}} \frac{\partial H_y}{\partial z} \tag{10}$$

$$E_z = \frac{-\beta}{\omega\varepsilon_0 \varepsilon_{NL}} H_y \tag{11}$$

Let us compute the first integral of (8):

$$\left[\frac{dH_y}{dz}\right]^2 - \gamma_{NL}^2 H_y + \frac{1}{2} k_0^2 \alpha' H_y^4 = S \tag{12}$$

By supposing $S \geq 0$ ($S$ is an integration constant), we will obtain:

$$\frac{dH_y}{dz} = \pm\sqrt{S + \gamma_{NL}^2 H_y - \frac{1}{2} k_0^2 \alpha' H_y^4} \tag{13}$$

It should be noted that for the case of $S \leq 0$, the expressions are similar. The equation of (13) has a familiar solution presented by Jacobi elliptic function (denoted by "cn") [58]:

$$H_y(z) = A_{NL} \, cn\big(q[(z-z_0) + z_{oc}], r\big) \tag{14}$$

where

$$q = \sqrt[4]{\gamma_{NL}^4 + 2S k_0^2 \alpha'} \tag{15}$$

$$r = \frac{q^2 + \gamma_{NL}^2}{2q^2} \tag{16}$$

$$A_{NL} = \frac{1}{k_0} \sqrt{\frac{q^2 + \gamma_{NL}^2}{\alpha'}} \tag{17}$$

Are used in (14). For the studied SPP wave here, as $z \to \infty$, $H_y \to 0$ and $\partial H_y/\partial z \to 0$ so that S=0. Thus, from the equations (15) and (16), we conclude that r=1. So, the above solution is written for $S = 0, r = 1$:

$$H_y(z) = \frac{\gamma_{NL}}{k_0} \sqrt{\frac{2}{\alpha'}} \, sech\big(\gamma_{NL}(z-z_0)\big) = A_{NL} \, sech\big(\gamma_{NL}(z-z_0)\big) \tag{18}$$

Now, we should write and solve Maxwell's equations inside LiF. LiF is a polar dielectric with the following permittivity [43]:

$$\varepsilon_{LiF}(\omega) = \varepsilon_{\infty}\left(1 - \frac{\omega_{LO}^2 - \omega_{TO}^2}{\omega^2 - \omega_{TO}^2 - j\omega\gamma}\right) \quad (19)$$

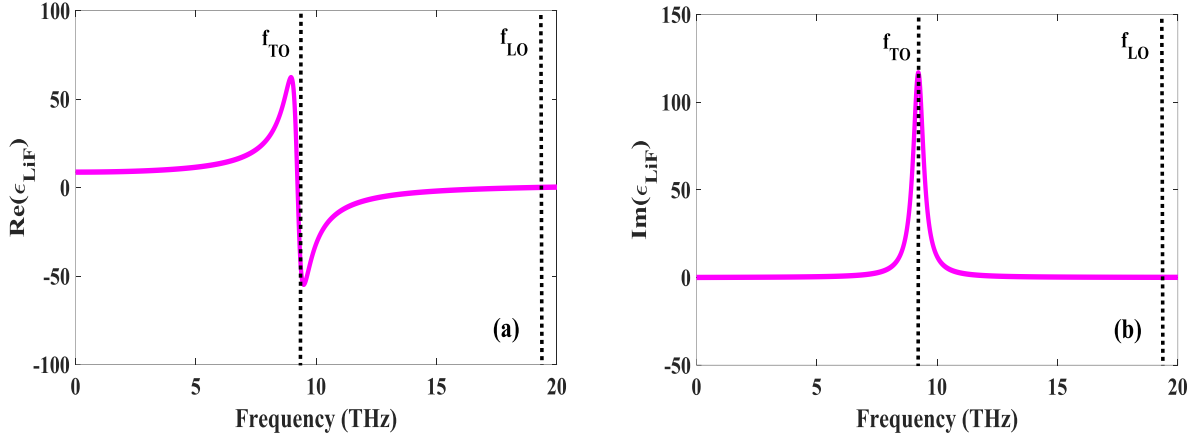

**Fig. 2.** The permittivity of LiF versus frequency: **(a)** Real part, **(b)** Imaginary part. In this figure, $f_{TO}, f_{LO}$ represent transverse and longitudinal optical frequencies, respectively.

Where $\varepsilon_{\infty} = 2.027, \omega_{TO} = 2\pi f_{TO}, \omega_{LO} = 2\pi f_{LO}, f_{TO} = 9.22\ THz, f_{LO} = 19.1\ THz, \Upsilon = 2\pi \times 0.527\ THz$ and $f_{TO}, f_{LO}, \Upsilon$ are the transverse optical frequency, longitudinal optical frequency, and the damping factor, respectively [43]. Fig. 2 shows the permittivity of LiF versus frequency. As seen in this figure, the real part of the permittivity has a resonance behavior near $f_{TO}$. In the Reststrahlen band ($f_{TO} < f < f_{LO}$), $Re[\varepsilon_{LiF}] < 0$ which means that this polar dielectric acts as a metal in this frequency band.

The second Maxwell's equation inside the LiF can be expressed as (the first one is similar to that inside the nonlinear medium) [59]:

$$\nabla \times \boldsymbol{H} = -j\omega\varepsilon_0\varepsilon_{LiF}\boldsymbol{E} \quad (20)$$

By doing some mathematical procedures, the wave equation for the TM mode inside the LiF layer is finally derived:

$$\frac{d^2 H_y}{dz^2} - \gamma_{LiF}^2 H_y = 0 \quad (21)$$

$$\gamma_{LiF}^2 = \beta^2 - k_0^2 \varepsilon_{LiF} \quad (22)$$

with the following transverse component of electric fields:

$$E_x = \frac{1}{j\omega\varepsilon_0\varepsilon_{LiF}}\frac{\partial H_y}{\partial z} \quad (23)$$

$$E_z = \frac{-\beta}{\omega\varepsilon_0\varepsilon_{LiF}}H_y \quad (24)$$

For the dielectric media including SiO₂ and Si layers, the above relation (i.e. relations (21)-(24)) can be utilized by simply substituting $\varepsilon_{LiF} \rightarrow \varepsilon_{Si}\ or\ \varepsilon_{SiO_2}$. Now, let us solve the propagating TM mode in the whole structure for various regions. The magnetic component of this wave is written as:

$$H_y(z) = \begin{cases} A_{NL}\ \text{sech}(\gamma_{NL}(z+z_1)) & z > 0 \\ Be^{-j\gamma_{LiF}z} + Ce^{+j\gamma_{LiF}z} & -t < z < 0 \\ De^{-j\gamma_{SiO_2}z} + Ee^{+j\gamma_{SiO_2}z} & -(t+d) < z < -t \\ Fe^{+\gamma_{Si}z} & z < -(t+d) \end{cases} \quad (25)$$

The unknown coefficients of $B, C, D, E, F$ in (25) should be determined by applying boundary conditions and $A_{NL}$ is given in (18). The $E_x$ component inside each layer can be obtained by (10) and (23),

$$E_x(z) = \frac{-1}{\omega\varepsilon_0}\begin{cases} \left(-jA_{NL}\dfrac{\gamma_{NL}}{\varepsilon_{NL}}\right)\text{sech}(\gamma_{NL}(z+z_1))\times\tanh(\gamma_{NL}(z+z_1)) & z > 0 \\ \dfrac{\gamma_{LiF}}{\varepsilon_{LiF}}\left(Be^{-j\gamma_{LiF}z} - Ce^{+j\gamma_{LiF}z}\right) & -t < z < 0 \\ \dfrac{\gamma_{SiO_2}}{\varepsilon_{SiO_2}}\left(De^{-j\gamma_{SiO_2}z} - Ee^{+j\gamma_{SiO_2}z}\right) & -(t+d) < z < -t \\ -j\dfrac{\gamma_{Si}}{\varepsilon_{Si}}Fe^{+\gamma_{Si}z} & z < -(t+d) \end{cases} \quad (26)$$

By applying the following boundary conditions,

$$E_{x,2} = E_{x,1} = E_x \quad (27)$$

$$H_{y,2} - H_{y,1} = \begin{cases} \sigma E_x & \text{if } z = 0 \\ 0 & \text{otherwise} \end{cases} \quad (28)$$

We finally obtain a system of nonlinear equations:

$$\begin{cases} a_1 f_{NL,z_1} + a_2 B - a_2 C = 0 \\ a_3 B + a_4 C + a_5 D + a_6 E = 0 \\ a_7 D + a_8 E + a_9 F = 0 \\ a_{10} g_{NL,z_1} + a_{11} B + a_{12} C = 0 \\ a_{13} B + a_{14} C + a_{15} D + a_{16} E = 0 \\ a_{17} D + a_{18} E + a_{19} F = 0 \end{cases} \quad (29)$$

With the following coefficients,

$$a_1 = -jA_{NL}\frac{\gamma_{NL}}{\varepsilon_{NL}},\ a_2 = -\frac{\gamma_{LiF}}{\varepsilon_{LiF}},\ a_3 = \frac{\gamma_{LiF}}{\varepsilon_{LiF}}e^{+j\gamma_{LiF}t},\ a_4 = -\frac{\gamma_{LiF}}{\varepsilon_{LiF}}e^{-j\gamma_{LiF}t}$$

$$a_5 = -\frac{\gamma_{SiO_2}}{\varepsilon_{SiO_2}}e^{+j\gamma_{SiO_2}t},\ a_6 = \frac{\gamma_{SiO_2}}{\varepsilon_{SiO_2}}e^{-j\gamma_{SiO_2}t},\ a_7 = \frac{\gamma_{SiO_2}}{\varepsilon_{SiO_2}}e^{+j\gamma_{SiO_2}(t+d)},\ a_8 = -\frac{\gamma_{SiO_2}}{\varepsilon_{SiO_2}}e^{-j\gamma_{SiO_2}(t+d)}$$

$$a_9 = j\frac{\gamma_{Si}}{\varepsilon_{Si}}e^{-\gamma_{Si}(t+d)},\ a_{10} = A_{NL},\ a_{11} = \sigma\frac{\gamma_{LiF}}{\varepsilon_{LiF}} - 1,\ a_{12} = -\sigma\frac{\gamma_{LiF}}{\varepsilon_{LiF}} - 1 \quad (30)$$

$$a_{13} = e^{+j\gamma_{LiF}t},\ a_{14} = e^{-j\gamma_{LiF}t},\ a_{15} = -e^{+j\gamma_{SiO_2}t},\ a_{16} = -e^{-j\gamma_{SiO_2}t}$$

$$a_{17} = e^{+j\gamma_{SiO_2}(t+d)},\ a_{18} = e^{-j\gamma_{SiO_2}(t+d)},\ a_{19} = -e^{-\gamma_{Si}(t+d)}$$

In (29), $f_{NL,z_1}, f_{NL,z_2}, g_{NL,z_1}, g_{NL,z_2}$ are defined by the following nonlinear functions:

$$f_{NL,z_1} = \text{sech}(\gamma_{NL} z_1) \times \tanh(\gamma_{NL} z_1) \tag{31}$$

$$g_{NL,z_1} = \text{sech}(\gamma_{NL} z_1) \tag{32}$$

By doing some complicated mathematical procedures, we finally achieve the following dispersion relation:

$$-j\frac{\gamma_{NL}}{\varepsilon_{NL}}\tanh(\gamma_{NL}z_1) = \frac{\begin{bmatrix} D_1(\omega).\cos(\gamma_{SiO_2}d).\cos(\gamma_{LiF}t) + D_2(\omega).\sin(\gamma_{SiO_2}d).\cos(\gamma_{LiF}t) + \\ D_3(\omega).\cos(\gamma_{SiO_2}d).\sin(\gamma_{LiF}t) + D_4(\omega).\sin(\gamma_{SiO_2}d).\sin(\gamma_{LiF}t) \end{bmatrix}}{\begin{bmatrix} D_5(\omega).\cos(\gamma_{SiO_2}d).\cos(\gamma_{LiF}t) + D_6(\omega).\sin(\gamma_{SiO_2}d).\cos(\gamma_{LiF}t) + \\ D_7(\omega).\cos(\gamma_{SiO_2}d).\sin(\gamma_{LiF}t) + D_8(\omega).\sin(\gamma_{SiO_2}d).\sin(\gamma_{LiF}t) \end{bmatrix}} \tag{33}$$

where

$$D_1(\omega) = -\frac{\gamma_{SiO_2}}{\varepsilon_{SiO_2}}\frac{\varepsilon_{LiF}}{\gamma_{LiF}} \tag{34}$$

$$D_2(\omega) = \frac{\gamma_{Si}}{\varepsilon_{Si}}\frac{\varepsilon_{LiF}}{\gamma_{LiF}} \tag{35}$$

$$D_3(\omega) = -j \tag{36}$$

$$D_4(\omega) = j\frac{\gamma_{Si}}{\varepsilon_{Si}}\frac{\varepsilon_{SiO_2}}{\gamma_{SiO_2}} \tag{37}$$

$$D_5(\omega) = \frac{\varepsilon_{LiF}}{\gamma_{LiF}}\left(1 + \sigma\frac{\gamma_{SiO_2}}{\varepsilon_{SiO_2}}\right) \tag{38}$$

$$D_6(\omega) = -\frac{\gamma_{Si}}{\varepsilon_{Si}}\frac{\varepsilon_{LiF}}{\gamma_{LiF}}\left(\sigma + \frac{\varepsilon_{SiO_2}}{\gamma_{SiO_2}}\right) \tag{39}$$

$$D_7(\omega) = j\left(\sigma - \frac{\gamma_{SiO_2}}{\varepsilon_{SiO_2}}\left(\frac{\varepsilon_{LiF}}{\gamma_{LiF}}\right)^2\right) \tag{40}$$

$$D_8(\omega) = j\frac{\gamma_{Si}}{\varepsilon_{Si}}\left(\left(\frac{\varepsilon_{LiF}}{\gamma_{LiF}}\right)^2 - \sigma\frac{\varepsilon_{SiO_2}}{\gamma_{SiO_2}}\right) \tag{41}$$

are defined in (33). The effective index, the propagation length, and the FOM can be defined as follows:

$$N_{eff} = \text{Re}[\beta] \tag{42}$$

$$L_{prop} = \frac{1}{2\,\text{Im}[\beta]} \tag{43}$$

$$FOM = \frac{\text{Re}[\beta]}{4\pi\,\text{Im}[\beta]} = \frac{N_{eff}.L_{prop}}{2\pi} \tag{44}$$

Now, our analytical mode for the proposed structure given in Fig. 1 is completed. Before embarking to investigate the behavior of HSP$^3$ in the heterostructure, let us consider some special cases to check the applicability of our general dispersion relation given in (33):

i) *Conventional SPPs on graphene sandwiched between two dielectric layers* ($\acute{\alpha} = 0, \varepsilon_{NL} = \varepsilon_L, t \to \infty$): In this case, the following well-known dispersion relation can be derived from (33) [60]:

$$\frac{\varepsilon_L}{\sqrt{\beta^2 - k_0^2 \varepsilon_L}} + \frac{\varepsilon_{LiF}}{\sqrt{\beta^2 - k_0^2 \varepsilon_{LiF}}} = -j\frac{\sigma}{\omega\varepsilon_0} \quad (45)$$

ii) *A multilayer system without graphene and LiF layers* ($\acute{\alpha} = 0, \varepsilon_{NL} = \varepsilon_L, \sigma = 0, t = 0$): In this case, the familiar dispersion relation for the three-layer optical waveguide is obtained by (33) [61]:

$$\tanh\left(d\sqrt{\beta^2 - k_0^2 \varepsilon_{SiO_2}}\right) = \frac{\dfrac{\varepsilon_{SiO_2}\gamma_L}{\varepsilon_L \gamma_{SiO_2}} + \dfrac{\varepsilon_{SiO_2}\gamma_{Si}}{\varepsilon_{Si}\gamma_{SiO_2}}}{1 + \dfrac{\varepsilon_{SiO_2}\gamma_L}{\varepsilon_L \gamma_{SiO_2}} \cdot \dfrac{\varepsilon_{SiO_2}\gamma_L}{\varepsilon_L \gamma_{SiO_2}}} \quad (46)$$

iii) *Air-graphene-SiO$_2$-Si multilayer structure* ($\acute{\alpha} = 0, \varepsilon_{NL} = \varepsilon_{air}, t = 0$): In this case, the general dispersion relation of (33) is converted to the following relation [62]:

$$\exp(-2\gamma_{SiO_2} d) = \frac{1 + \dfrac{\varepsilon_{air}\gamma_{SiO_2}}{\varepsilon_{SiO_2}\gamma_{air}}\left(1 + j\dfrac{\sigma\gamma_{air}}{\omega\varepsilon_0}\right)}{1 - \dfrac{\varepsilon_{air}\gamma_{SiO_2}}{\varepsilon_{SiO_2}\gamma_{air}}\left(1 + j\dfrac{\sigma\gamma_{air}}{\omega\varepsilon_0}\right)} \cdot \frac{1 + \dfrac{\varepsilon_{Si}\gamma_{SiO_2}}{\varepsilon_{SiO_2}\gamma_{Si}}}{1 - \dfrac{\varepsilon_{Si}\gamma_{SiO_2}}{\varepsilon_{SiO_2}\gamma_{Si}}} \quad (47)$$

## 3. Results and Discussions

This section investigates the analytical results obtained by our mathematical relations in the previous section. To simulate the proposed structure, the graphene parameters are supposed to be: $\mu_c = 0.2\ eV, \tau = 0.2\ ps, T = 300\ K, \Delta = 1 nm$. The permittivity of Si and SiO$_2$ layers are assumed to be $\varepsilon_{Si} = 11.9, \varepsilon_{SiO_2} = 2.09$, respectively. The parameters of the nonlinear medium are chosen: $\alpha = 6.4 \times 10^{-12} m^2 V^{-2}, \varepsilon_L = 2.405, \alpha'|H_{y,0}(0)|^2 = 0.5$ [57]. The parameters of the LiF layer are given before. Moreover, the configuration parameters are supposed to be $t = 10 nm,, d = 200 nm$.

Fig. 3 illustrates the effective index, the propagation length, and the FOM of the propagating HSP$^3$ as a function of frequency. As observed in Fig. 3(a), the effective index or the propagation constant is affected by plasmon-phonon coupling, where the phonon oscillations of LiF inside the Reststrahlen band are dominant. Furthermore, one can see from this figure that the backward waves with the negative slope of dispersion are excited in two frequency regions: 1) near the transverse optical frequency (9.22 THz), 2) inside the Reststrahlen band ($17.1\ THz < f < 18.8\ THz$). These waves are similar to the ones supported by metal-insulator-metal structures [63, 64] and originate from the phononic behavior of LiF material. It can be observed from fig. 3(b) that the propagation length decreases as the frequency increases and it has low levels of value in the Reststrahlen band.

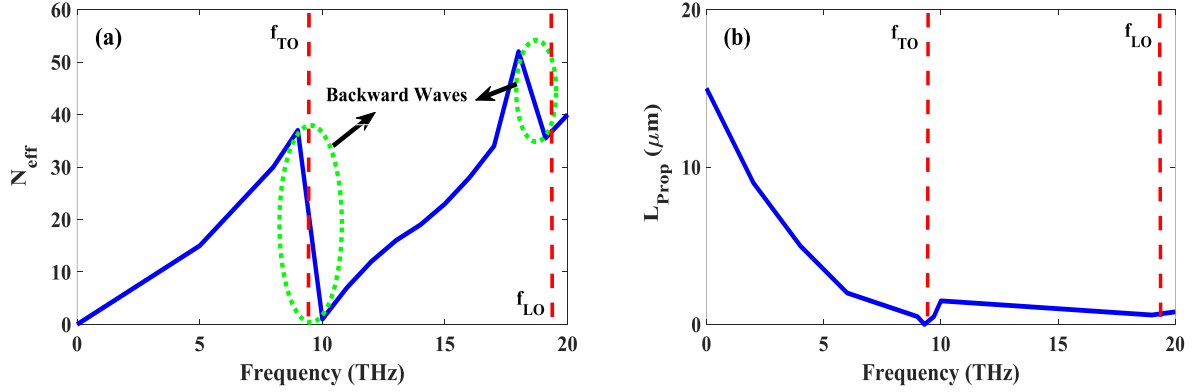

**Fig. 3. (a)** Effective Index, **(b)** Propagation length of the propagating SP$^3$s as a function of frequency for the proposed waveguide. In these diagrams, the thickness of the LiF layer is supposed to be $t = 10\ nm$. Other design parameters remained fixed.

Let us consider what happens if we increase the thickness of the LiF layer to be $t = 100\ nm$. The characteristic properties of the proposed structure have been depicted in Fig. 4 for $t = 100\ nm$. We see from Fig. 4(a) that there are no backward waves near the transverse optical frequency (9.22 THz) and they are turned to forward waves, compared to Fig. 3 (a) for $t = 10\ nm$. However, in the Reststrahlen band, the backward waves propagate, similar to $t = 10\ nm$. The propagation length of the propagating HSP$^3$ has been enhanced as the thickness of LiF increases. It is observable from fig. 4(c) that the maximum value of FOM in the Reststrahlen band reaches 5.8 at the frequency of 14.3 THz while it has a maximum value of 11.9 at the frequency of 9.22 THz outside the Reststrahlen band.

Fig. 5 shows the influence of the nonlinearity on the propagation feature of HSP$^3$ as a function of frequency. The thicknesses of LiF and SiO$_2$ layers are supposed to be $t = 100\ nm, d = 200\ nm$, respectively. The chemical potential of graphene is $\mu_c = 0.2\ ev$ and the relaxation time is assumed to be $\tau = 0.2\ ps$. As seen in this figure, the FOM of HSP$^3$ increases in the vicinity of the transverse optical frequency (9.22 THz). It reaches 24.5 at the frequency of 9.22 THz, which is desirable FOM for many potential applications in the far-infrared region. In the phononic region, its value has a maximum of around 14.3 THz. Furthermore, increasing nonlinearity will increase the FOM inside this band. However, for the backward waves in the Reststrahlen band, the nonlinearity has a negligible effect on FOM. From Fig. 5, it seems that the nonlinear coefficient can effectively enhance the propagation characteristics of HSP$^3$.

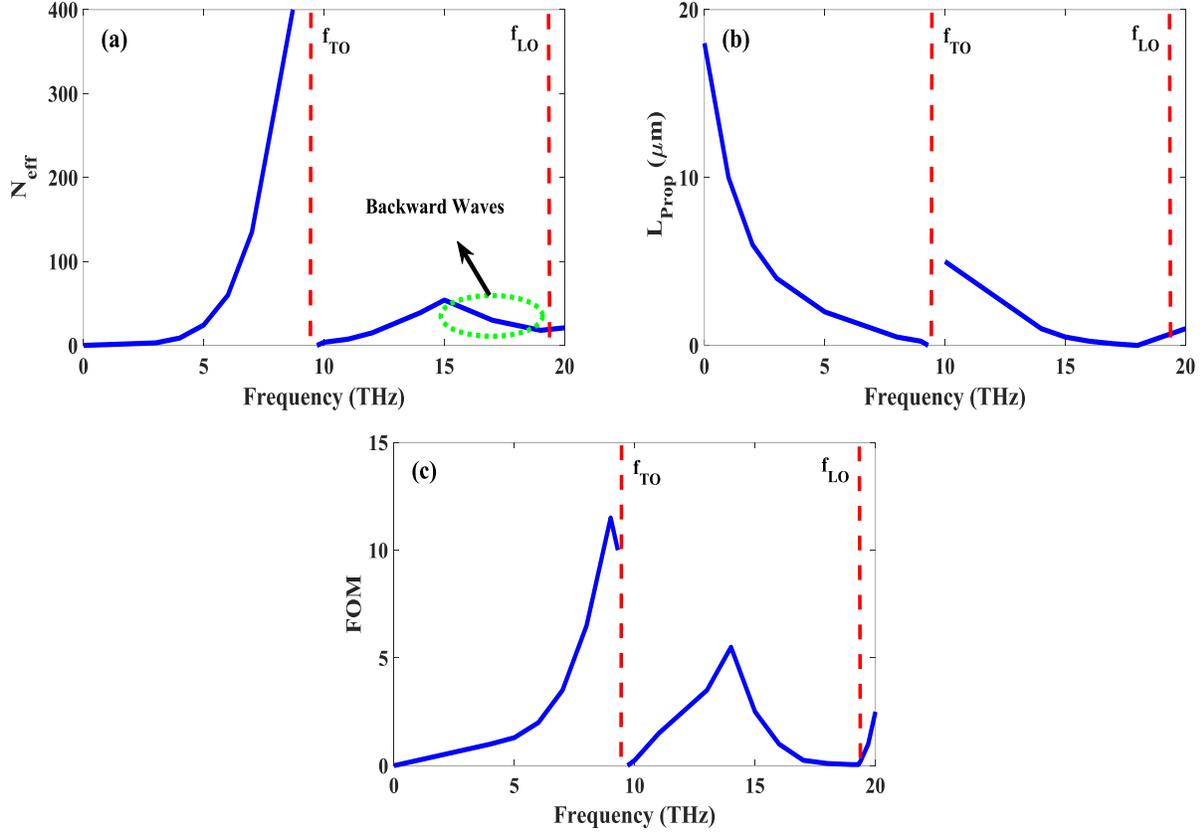

**Fig. 4.** (a) Effective Index, (b) Propagation length, and (c) FOM of the propagating SP$^3$s as a function of frequency for the proposed waveguide. In these diagrams, the thickness of the LiF layer is supposed to be $t = 100\ nm$. Other design parameters remained fixed.

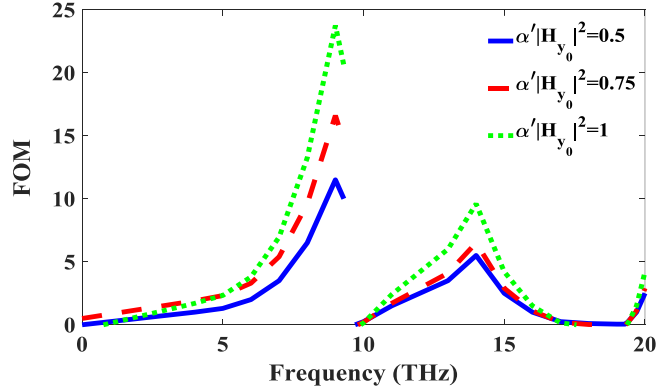

**Fig. 5.** FOM of the proposed waveguide as a function of frequency for various values of the nonlinear coefficient ($\alpha'|H_{y,0}(0)|^2 = 0.5, 0.75, 1$). The thicknesses of LiF and SiO$_2$ layers are supposed to be $t = 100\ nm, d = 200\ nm$, respectively. The chemical potential of graphene is $\mu_c = 0.2\ ev$ and the relaxation time is assumed to be $\tau = 0.2\ ps$.

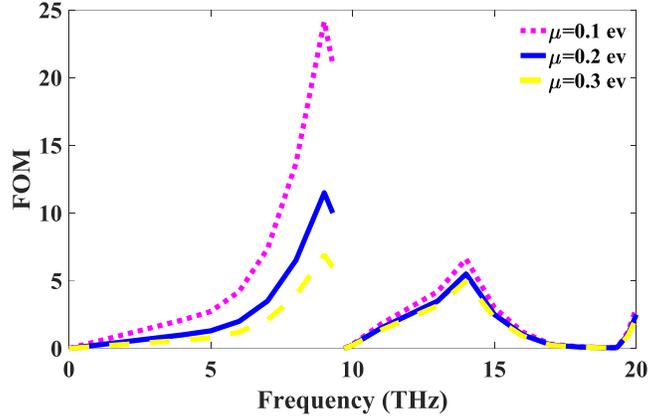

**Fig. 6.** FOM of the proposed waveguide as a function of frequency for various values of the chemical potential ($\mu_c = 0.1, 0.2, 0.3\ ev$). The thicknesses of LiF and SiO$_2$ layers are supposed to be $t = 100\ nm, d = 200\ nm$, respectively. The nonlinear coefficient is $\alpha'|H_{y,0}(0)|^2 = 0.5$ and the relaxation time is assumed to be $\tau = 0.2\ ps$.

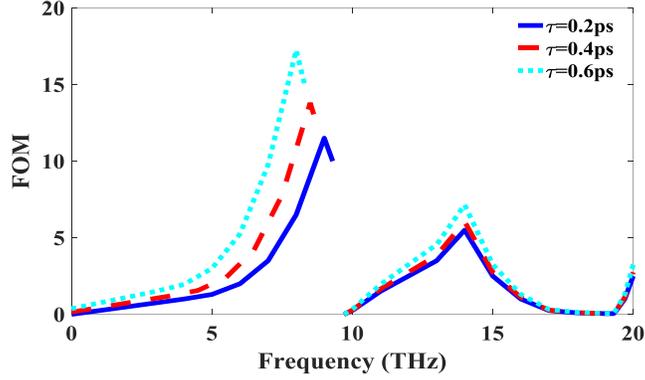

**Fig. 7.** FOM of the proposed waveguide as a function of frequency for various values of the relaxation time ($\tau = 0.2, 0.4, 0.6\ ps$). The thicknesses of LiF and SiO$_2$ layers are supposed to be $t = 100\ nm, d = 200\ nm$, respectively. The nonlinear coefficient is $\alpha'|H_{y,0}(0)|^2 = 0.5$ and the chemical potential is assumed to be $\mu_c = 0.2\ ev$.

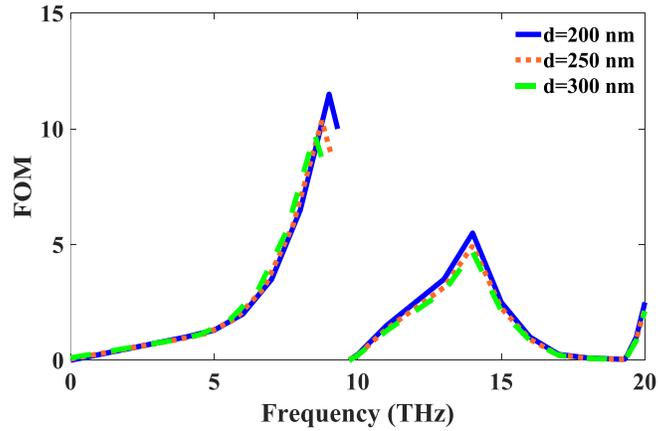

**Fig. 8.** FOM of the proposed waveguide as a function of frequency for various values of the SiO$_2$ thickness ($d = 200, 250, 300\ nm$). The thickness of the LiF layer is supposed to be $t = 100\ nm$. The chemical potential of graphene is $\mu_c = 0.2\ ev$ and the relaxation time is assumed to be $\tau = 0.2\ ps$.

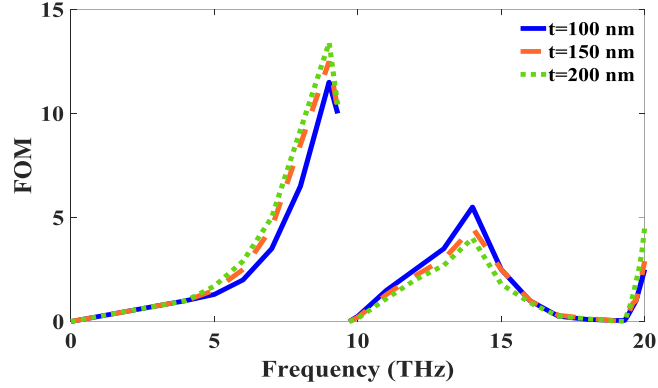

**Fig. 9.** FOM of the proposed waveguide as a function of frequency for various values of the LiF thickness ($t = 100, 150, 200\ nm$). The thickness of the SiO$_2$ layer is supposed to be $d = 200\ nm$. The chemical potential of graphene is $\mu_c = 0.2\ ev$ and the relaxation time is assumed to be $\tau = 0.2\ ps$.

As mentioned before, our proposed heterostructure is tunable and its propagating properties can be varied via the graphene parameters. Fig. 6 clearly demonstrates this, where FOM has been depicted as a function of frequency for various values of the chemical potential. In this figure, other parameters have remained fixed. Outside the Reststrahlen band, as the chemical potential increases, FOM decreases due to the increment of wave number. Moreover, as observed in this figure, changing the chemical potential of the graphene layer has a negligible effect on FOM inside the Reststrahlen band, as expected. As discussed before, the plasmonic feature of plasmon-phonon coupling is dominant outside the Reststrahlen band while the phononic behavior of HSP$^3$ dominates inside the Reststrahlen band. For the chemical potential of 0.1 eV, FOM reaches 24.9 at the frequency of 9.22 THz. One of the main parameters of a graphene sheet is the phenomenological relaxation time ($\tau$) or the electron mobility. In Fig. 7, we have illustrated the FOM for various values of the relaxation time. Similar to Fig. 6, in the Reststrahlen band, the effect of this parameter on FOM is negligible while it changes FOM outside this band.

To achieve the best performance by the suitable design of the structure, the configuration parameters should be chosen more precisely. Fig. 8 shows the FOM as a function of frequency for various values of SiO$_2$ thickness (d). As seen in this figure, the thickness of the SiO$_2$ layer has a negligible effect on the FOM of propagating HSP$^3$. It happens because the concentration of HSP$^3$ mode is at the vicinity of the graphene-LiF border. Hence, our design is not so sensitive to the value of this parameter.

In Fig. 9, we present the effect of LiF thickness (t) on FOM for values of $t > 100\ nm$. As discussed before, increasing the LiF thickness from 10nm to 100nm turned the backward waves into forward waves outside the Reststrahlen band. However, as seen in Fig. 9, for values of $t > 100\ nm$, the changes in LiF thickness have a slight effect on FOM. Outside the Reststrahlen band, the FOM of HSP$^3$ increases slightly with the increment of LiF thickness. The ability to tune the propagation characteristics of HSP$^3$ via the chemical potential and the nonlinearity can make the proposed structure a promising candidate for highly potential applications such as waveguides, sensors, and switches in the infrared region.

In this paper, we presented an analytical model containing closed-form relations for the propagation parameters of the proposed heterostructure. Our analytical model has two interesting advantages: 1) the speed of calculation is high because there is no numerical method faster than an analytical model, 2) the accuracy of the proposed method is also high compared to other computational methods. However, our model has some limitations and drawbacks: it considers only TM propagating mode and thus cannot be utilized for TE or hybrid modes. Furthermore, the propagating direction is supposed to be in the x-direction, hence the analytical model cannot cover the propagating waves in other directions.

As a final point, it should be noted that this article is an analytical work, and discussing the fabrication process of the proposed heterostructure is outside the scope of this paper. However, the following fabrication process is suggested: first, the graphene sheet is grown on a nonlinear substrate by chemical vapor deposition (CVD). Then, the LiF layer is deposited on the grown graphene by ion irradiation. Finally, $SiO_2$ and Si layers are placed on the LiF layer.

## 4. Conclusion

This paper studied the propagating characteristics of a new nano-structure, constituting graphene-LiF-Nonlinear layers in the far-infrared frequencies. An analytical model for the proposed heterostructure was presented, embarking on Maxwell's equations, deriving a wave equation for each layer, and then applying the boundary conditions at the borders. In this article, a complicated dispersion relation was derived for the proposed structure. The analytical results showed that a high value of the figure of merit (FOM), i.e. FOM=24.5, could be obtained at the frequency of 9.22 THz. Furthermore, it was shown that the propagating features of our heterostructure for Hybrid Surface Plasmon Phonon Polaritons in three regions, i.e. inside, below, and above the Reststrahlen band, could be effectively tuned by the chemical potential and the nonlinear coefficient. Our proposed structure can be a promising candidate for multi-functional applications such as waveguides, absorbers, and sensors in the THz region.